\documentclass[pra,aps,preprint,superscriptaddress,showpacs,nofootinbib]{revtex4-1}

\usepackage[dvips]{graphicx}
\usepackage{amsmath,amssymb,amsthm,mathrsfs,amsfonts,dsfont}
\usepackage{subfigure, epsfig}
\usepackage{braket}
\usepackage{bm}
\usepackage{enumerate}
\usepackage{color}

\newcommand{\ICFOAddress}{ICFO -- Institut de Ciencies Fotoniques, The Barcelona Institute of Science and Technology, 08860 Castelldefels (Barcelona), Spain} 
\newcommand{\ICREAAddress}{ICREA -- Instituci\'{o} Catalana de Recerca i Estudis Avan\c{c}ats, 08010 Barcelona, Spain} 

\newcommand{\USTCAddress}{Shanghai Branch, National Laboratory for Physical Sciences at Microscale and Department of Modern Physics, University of Science and Technology of China, Shanghai 201315, P.~R.~China} 
\newcommand{\QPQIAddress}{Shanghai Branch, CAS Center for Excellence and Synergetic Innovation Center in Quantum Information and Quantum Physics, University of Science and Technology of China, Shanghai 201315, P.~R.~China}
\newcommand{\TsinghuaAddress}{Center for Quantum Information, Institute for Interdisciplinary Information Sciences, Tsinghua University, Beijing 100084, P.~R.~China}
\newcommand{\SIMITAddress}{State Key Laboratory of Functional Materials for Informatics, Shanghai Institute of Microsystem and Information Technology, Chinese Academy of Sciences, Shanghai 200050, P.~R.~China}
\newcommand{\PMOAddress}{Purple Mountain Observatory and Key Laboratory of Radio Astronomy, Chinese Academy of Sciences, 2 West Beijing Road, Nanjing, Jiangsu 210008, P.~R.~China}

\begin{document}

\title{Experimental test of measurement dependent local Bell inequality with human free will}

\author{Yang Liu}
\affiliation{\USTCAddress}
\affiliation{\QPQIAddress}

\author{Xiao Yuan}
\affiliation{\USTCAddress}
\affiliation{\QPQIAddress}
\affiliation{\TsinghuaAddress}

\author{Cheng Wu}
\affiliation{\USTCAddress}
\affiliation{\QPQIAddress}

\author{Weijun Zhang}
\affiliation{\SIMITAddress}

\author{Jian-Yu Guan}
\affiliation{\USTCAddress}
\affiliation{\QPQIAddress}

\author{Jiaqiang Zhong}
\affiliation{\PMOAddress}

\author{Hao Li}
\affiliation{\SIMITAddress}

\author{Ming-Han Li}
\affiliation{\USTCAddress}
\affiliation{\QPQIAddress}

\author{Carlos Abell\'{a}n}
\affiliation{\ICFOAddress}
 
\author{Morgan W. Mitchell}
\affiliation{\ICFOAddress}
\affiliation{\ICREAAddress}

\author{Sheng-Cai Shi}
\affiliation{\PMOAddress}

\author{Jingyun Fan}
\affiliation{\USTCAddress}
\affiliation{\QPQIAddress}

\author{Lixing You}
\affiliation{\SIMITAddress}

\author{Zhen Wang}
\affiliation{\SIMITAddress}

\author{Xiongfeng Ma}
\affiliation{\TsinghuaAddress}

\author{Qiang Zhang}
\affiliation{\USTCAddress}
\affiliation{\QPQIAddress}

\author{Jian-Wei Pan}
\affiliation{\USTCAddress}
\affiliation{\QPQIAddress}

\begin{abstract}
A Bell test can rule out local realistic models, and has potential applications in communications and information tasks.  For example, a Bell inequality violation can certify the presence of intrinsic randomness in measurement outcomes, which then can be used to generate unconditional randomness.  A Bell test requires, however, measurements that are chosen independently of other physical variables in the test, as would be the case if the measurement settings were themselves unconditionally random. This situation seems to create a ``bootstrapping problem'' that was recently addressed in The BIG Bell Test, a collection of Bell tests and related tests using human setting choices. Here we report in detail our experimental methods and results within the BIG Bell Test. We perform a experimental test of a special type of Bell inequality - the measurement dependent local inequality. With this inequality, even a small amount of measurement independence makes it possible to disprove local realistic models. The experiment uses human-generated random numbers to select the measurement settings in real time, and implements the measurement setting with space-like separation from the distant measurement. The experimental result shows a Bell inequality violation that cannot be explained by local hidden variable models with independence parameter (as defined in [P\"{u}tz et al. Phys. Rev. Lett. 113, 190402 (2014).] ) $l > 0.10 \pm 0.05$. This result quantifies the degree to which a hidden variable model would need to constrain human choices, if it is to reproduce the experimental results. 
\end{abstract}

\maketitle

A Bell test \cite{bell} is designed to rule out local hidden variable models (LHVMs) \cite{Einstein35}. By observing Bell inequality violations in experiments that faithfully reproduce the assumptions of Bell's theorem, we can demonstrate that the underlying physical process cannot be explained with LHVMs. In quantum information processing, Bell tests provide device independent advantages in a variety of tasks, such as quantum key distribution \cite{Mayers98, acin06,masanes2011secure,Vazirani14}, randomness amplification \cite{colbeck2012free, gallego2013full, Dhara14,brandao2016realistic,Ramanathan16} and expansion \cite{Colbeck11, Fehr13, Pironio13, vazirani2012certifiable}, entanglement quantification \cite{Moroder13}, and dimension witness \cite{Brunner08}.

Focusing on the bipartite scenario, a Bell test involves two space-like separated participants Alice and Bob, who implement measurements with settings, or ``inputs'' $x$ and $y$ and generate outputs $a$ and $b$, respectively. The Bell inequality is defined by a linear combination of the probability distribution $P(ab|xy)$ according to
\begin{equation}\label{Eq:Bellinequality}
  J = \sum_{abxy} c_{abxy}P(ab|xy) \leq J_C.
\end{equation}
Here, $J_C$ is the classical upper bound with LHVMs. With quantum mechanics, a Bell value larger than $J_C$ may be achieved. In such a case, we call it a violation of the Bell inequality.
In device independent tasks, the quantum advantage can be certified solely by the Bell value $J$ and the certification can be independent of the implementation of the devices used for state preparation and measurement. 
For instance, when the Bell inequality is violated, the output cannot be completely predicted, hence the entropy of the output is positive \cite{Nath2017}.  A complete random bits can be achieved using randomness extraction with the output from several rounds.

In practice, a faithful implementation of a Bell test is conditioned on closing three major loopholes.
The locality loophole: the generation of Alice's input $a$ should be space-like separated from Bob's measurement of $y$, and likewise for $b$ and $x$. If this condition is not satisfied, a Bell inequality can be violated even with LHVMs by signaling the inputs.
The efficiency loophole: the efficiency must be higher than a threshold to ensure the violation. If the realized efficiency is lower than the threshold, the violation cannot be observed without post-selecting the outputs.
The so-called ``freedom-of-choice'' loophole: the inputs should not be influenced by the hidden variables in the LHVM. 
Clearly, if the LHVM can determine the inputs any probability distribution $P(ab|xy)$ can be realized with a LHVM. This motivates efforts to choose the inputs randomly, or at least in a manner that cannot be controlled by a LHVM.

The history of testing Bell's inequality is the history trying to close all the possible loopholes, especially the locality and efficiency loopholes. It was only recently, these loopholes have been closed simultaneously in Bell tests using physical random number generators \cite{hensen2015loophole, Shalm15, Giustina15, Rosenfeld2017}. 
Nevertheless, the freedom-of-choice loophole cannot be closed perfectly, as we can never unconditionally certify the randomness without a faithful Bell test. This seems to create a ``bootstrapping problem", in which unconditional randomness is required in order to produce unconditional randomness. In principle, it not possible to rule out super-determinism, the philosophical position that the universe is completely deterministic.

When considering the practical case of certifying randomness in the presence of classical noise or an adversary, we can still assume the possibility that the input is random with respect to the measurement devices.
In experiment, well-calibrated quantum random number generators are used, with the assumption that their output values are independent of any prior events. In another words, the measurements performed in Bell tests are independent of the random inputs, named by measurement independence \cite{putz2014}. Photons from cosmic sources can be also considered as the random input in both theory and experiment, by pushing measurement dependence constraint back into cosmic history \cite{Gallicchio14,Handsteiner17,cheng2016random}. Nevertheless, this randomness is guaranteed by certain physical models, which can be inaccurate and hence make the random input partially predictable.

Without relying on any physical model, it is a common belief that humans have free will, by which we mean humans can make choices that are not deterministic consequences of prior physical conditions.  Assuming that humans indeed have this capacity, the use of human choices in a Bell test makes it possible to circumvent the bootstrapping problem described above. At the same time, human choices are not perfectly unpredictable - sequences of such choices tend to show patterns even when the person is attempting to make random choices.  For example,  we measured the uniformity of the human generated random numbers input from 30th Nov. to 1st Dec, 2016 in the BIG Bell Test. Using the well-known statistical test suite NIST SP 800-22 ("the NIST tests"), we found that of the 14 different tests, the human input only passed two.  In fact, we can easily find  ``000000'' or ``010101'' patterns in the human random number file. In contrast, our quantum random number generators passed all the tests.

Such patterns by themselves do not open the freedom-of-choice loophole because it is possible for two variables, here $xy$ and $\lambda$, to be independent even if one of them, here $xy$, is biased and thus somewhat predictable. Nevertheless, it is interesting to consider the possibility that some of the predictability of human choices arises due to influence by the physical environment, which then might be correlated with the hidden variables. Indeed, John Bell himself discussed this possibility \cite{bell1987speakable}. If such influence were too strong, a LHVM could explain a Bell inequality violation through a limited freedom of choice.  Theoretical analysis has made precise these qualitative notions about partially predictable inputs in Bell tests \cite{Hall10,Koh12,Koh12,Pope13,putz14,Yuan15,Yuan15b}. 

In this work, we exploit human randomness generated via free will as the input of Bell tests. The human randomness is collected in a worldwide project---the BIG Bell Test \cite{BBT2018}, initiated by researchers from ICFO. Basically, the main idea of the BIG Bell Test is to apply human free will to a state-of-the-art physics experiment, the Bell test. 
In this letter, we employ two different Bell inequalities, closing the locality loophole, and analyzing the requirements that the human randomness should satisfy to guarantee faithful violations of the Bell tests.

\emph{Measurement dependence.}---In this section, we review the theory of Bell tests with imperfect inputs, i.e., \emph{measurement dependence}. Given LHVMs, the probability distribution can be represented as
\begin{equation}\label{Eq:LHVMs}
	P(ab|xy) = \sum_\lambda P(\lambda) P(a|x\lambda)P(b|y\lambda).
\end{equation}
Here, $\lambda$ is the predetermined strategy shared by Alice and Bob and $P(\lambda)$ is the probability of choosing $\lambda$. 
In contrast, the probability distribution generated by measuring a quantum state cannot, in general, be described by Eq.~\eqref{Eq:LHVMs}. 
Bell inequalities can be regarded as witnesses of a quantum probability distribution that separates it from probability distributions of Eq.~\eqref{Eq:LHVMs}.

\begin{figure}[bth]
\centering \resizebox{4cm}{!}{\includegraphics{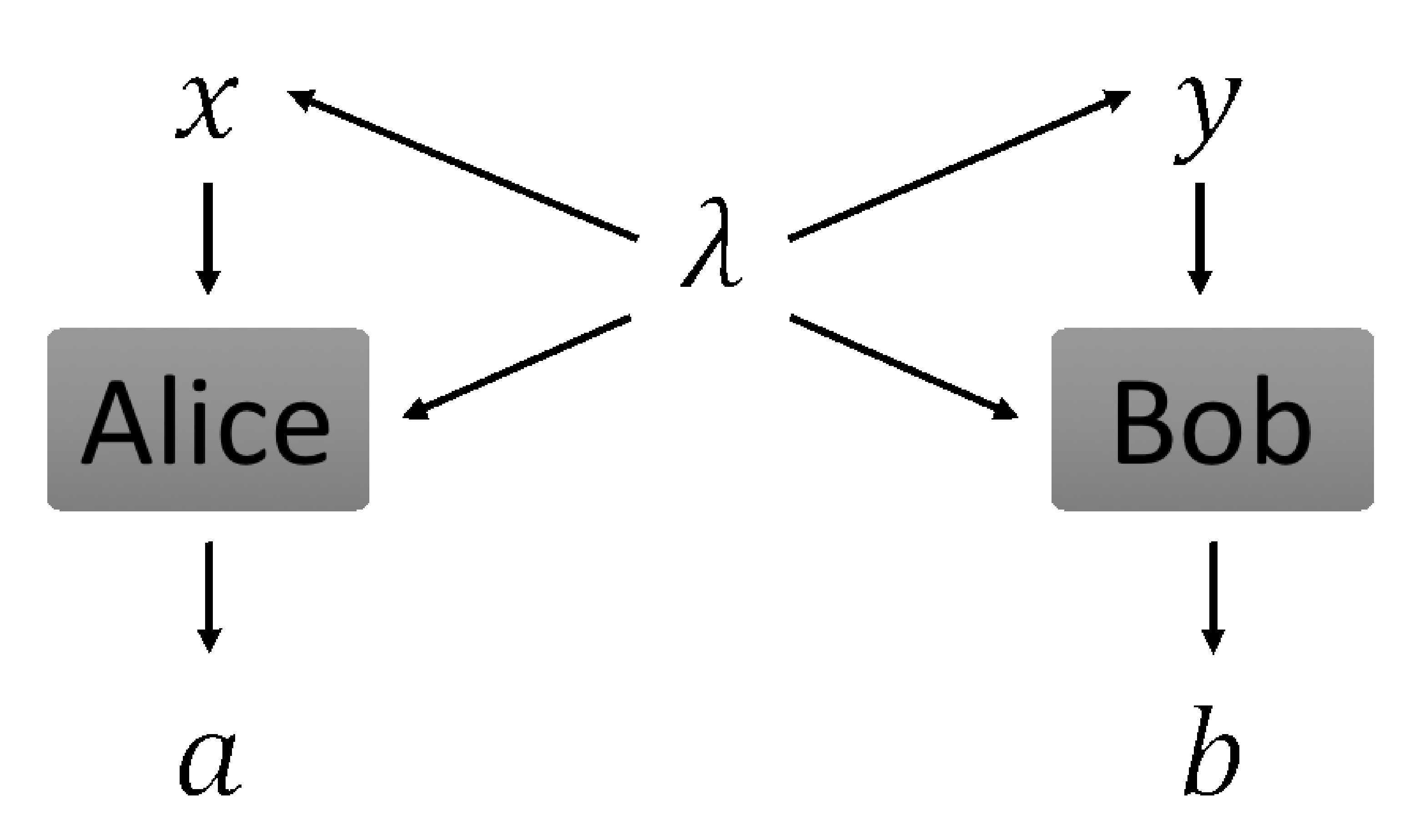}}
\caption{Bell test with imperfect input randomness.}
\label{Fig:MDL}
\end{figure}

In the ideal case, the inputs are assumed perfectly random. That is, the adversary does not have extra information of $x$ and $y$ better than blindly guessing. In practice, the input randomness can be imperfect and determined by some local hidden variable $\lambda$ according to $P(xy|\lambda)$. In this case, as shown in Fig.~\ref{Fig:MDL}, an adversary that has access to $\lambda$ can realize the measurement by exploiting such information and have a larger set of LHVMs with probability distributions of the form
\begin{equation}\label{Eq:MDL}
	P(abxy) = \sum_\lambda P(\lambda) P(xy|\lambda)P(a|x\lambda)P(b|y\lambda).
\end{equation}

To understand how $P(xy|\lambda)$ characterizes the input randomness, we consider the lower bounds of $P(xy|\lambda)$ defined as
\begin{equation}
\begin{aligned}
	l &= \min_{xy\lambda}P(xy|\lambda).
\end{aligned}
\end{equation}
Focusing on the case with binary inputs hereafter, we have $l\in[0,1/4]$.
We consider the two extreme cases. When the inputs are perfectly random, we have $P(xy|\lambda) = P(xy)$ and $l=1/4$. In this case, we recover the probability distribution given in Eq.~\eqref{Eq:LHVMs}. When the inputs are totally determined by $\lambda$, i.e., $(xy) = f(\lambda)$, the probability becomes $P(xy|\lambda) = \delta_{(xy),f(\lambda)}$, where $\delta$ is the Kronecker delta and we have $l=0$ . In such a case, it is possible to realize any probability distribution and the violation of a Bell inequality is possible with a LHVM.

From these two examples, we can infer that $l$ measures the input randomness. A smaller value of $l$ indicates that the input is less random or Alice and Bob have more information of the inputs. In the literature, the problem of Bell tests with imperfect inputs is named by {measurement dependence}. Contrarily, given $l$, when a probability distribution cannot be described by Eq.~\eqref{Eq:MDL}, we call it measurement dependent nonlocality. We now see that in a Bell test with human inputs, the assumption that humans have free will can be represented as $l>0$, implying at least partial measurement independence.  In the following, we consider two Bell inequalities and show how the Bell value is dependent on the input randomness parameter $l$.

\begin{figure*}[tbh]
\centering
\resizebox{13cm}{!}{\includegraphics{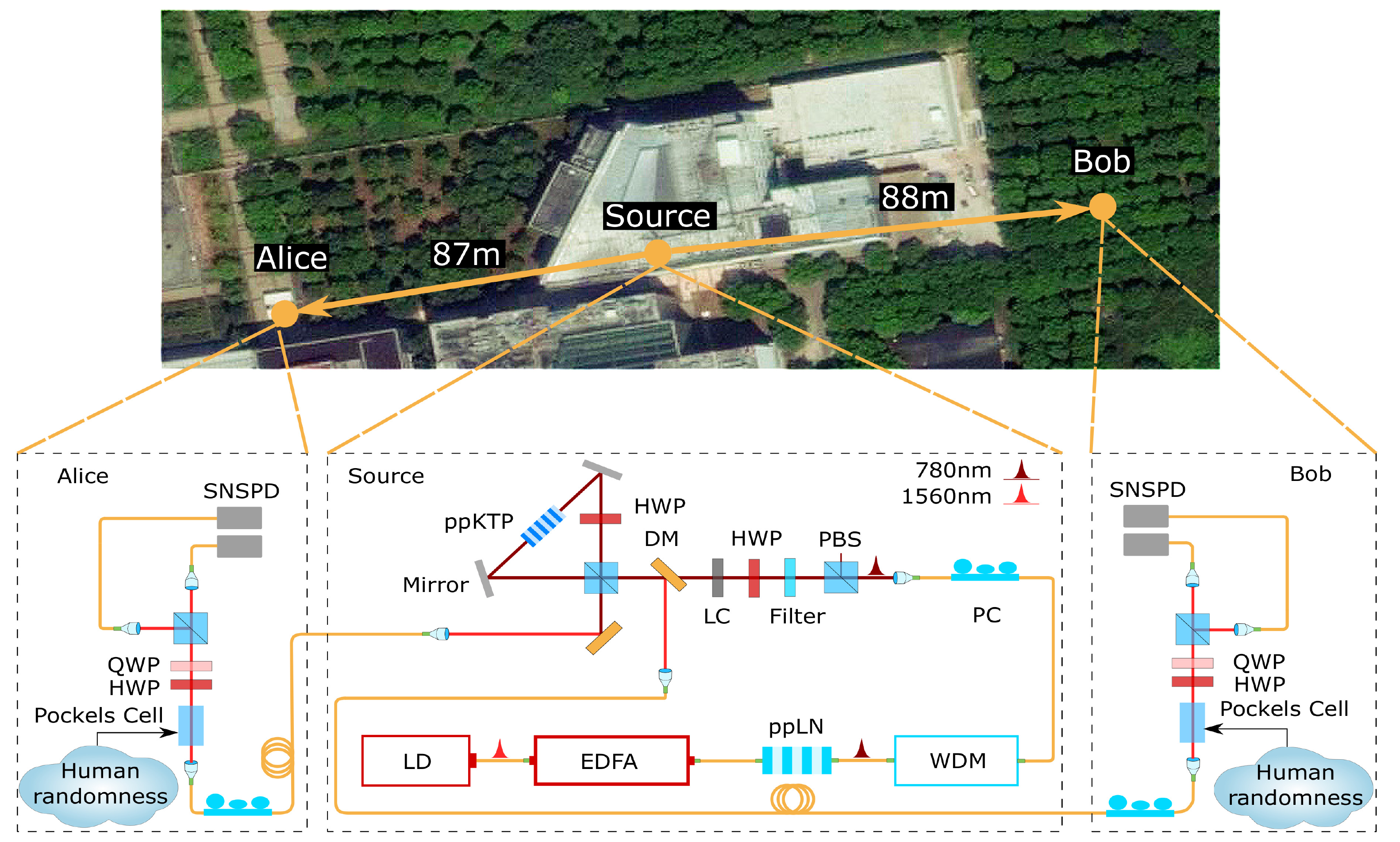}}
\caption{Bell test using imperfect input randomness. (a) The positions of the entanglement source, Alice's detection and Bob's detection. The distance between the source and Alice (Bob) is $87\pm2$ m ($88\pm2$ m). (b) Schematic setup of the Bell test. A distributed feedback (DFB) laser diode (LD) at $\lambda=1560$ nm is modulated to be 100 kHz with 10 ns width. The pulse is amplified with an erbium-doped fiber amplifier (EDFA) and then up converted to 780 nm via a second-harmonic generation (SHG) in an in-line periodically poled lithium niobate (PPLN) waveguide. The residual 1560 nm light is filtered with a wavelength-division multiplexer (WDM) and a filter. After adjusting the polarization using a half-wave plate (HWP) and a liquid crystal (LC), the 780 nm pump light is focused to a periodically poled potassium titanyl phosphate (PPKTP) crystal in a Sagnac setup to generate entangled pairs. A series of dichroic mirrors (DMs) are used to remove the residual pump light at 780 nm and fluorescence before the entangled pairs are collected. In Alice's and Bob's detection station, a polarization controller (PC), a quarter-wave plate (QWP), a HWP and a polarizing beam splitter (PBS) is used to align the reference frame. Random numbers control the Pockels cell to select the bases dynamically. Superconducting nanowire single-photon detectors (SNSPD) are used to detect the photons after PBS.}
\label{Fig:MDLSetup}
\end{figure*}

The coefficients of the Clauser-Horne-Shimony-Holt (CHSH) \cite{CHSH} are given by $c_{abxy} = (-1)^{a+b+xy}$.
Suppose the average probability $P(xy)=\sum_\lambda P(\lambda)P(xy|\lambda)$ equals $1/4$ and $P(ab|xy)$ is no-signaling \cite{prbox}, then the classical upper bound $J_C(l)$ with imperfect input randomness characterized by $l$ is \cite{putz14, Yuan15b}
\begin{equation}\label{Eq:BCHSH}
	J_C(l) = 4(1-2l).
\end{equation}
In experiment, the Bell inequality is violated only when the observed Bell value is larger than $J_C(l)$. To do so, a maximally entangled state is prepared
\begin{equation}
	\ket{\Psi^+}=(\ket{HV}+\ket{VH})/\sqrt{2},
\end{equation}
and measured in the following projecting bases
\begin{equation}\label{Eq:AngleCHSH}
\begin{aligned}
	{A_0(\theta)} = Z, \quad & {B_0(\theta)} = \frac{X-Z}{\sqrt{2}},\\
	{A_1(\theta)} = X, \quad &	{B_1(\theta)} = \frac{X+Z}{\sqrt{2}}.
\end{aligned}
\end{equation}
Here, $A_0$ and $A_1$ ($B_0$ and $B_1$) are the measurement bases of Alice (Bob) when the inputs are 0 and 1, respectively; $Z = \{\ket{H},\ket{V}\}$ and $X = \{(\ket{H}+\ket{V})/\sqrt{2},(\ket{H}-\ket{V})/\sqrt{2}\}$.

In this letter, we also consider another Bell inequality named the measurement dependent local (MDL) inequality proposed by P\"utz et al. \cite{putz14},
\begin{equation}\label{Eq:MDLin}
	lP(0000) -(1-3l)(P(0101) + P(1010) + P(0011)) \le 0.
\end{equation}
It is proven that when $l\le P(xy|\lambda)$ the probability distribution given in Eq.~\eqref{Eq:MDL} cannot violate such an inequality.
To violate this inequality with quantum settings, we need to prepare a non-maximally entangled state
\begin{equation}\label{Eq:Bellinequality}
 \ket{\Psi}=\frac{1}{\sqrt{3}}\left(\frac{\sqrt{5}-1}{2}\ket{HV}+\frac{\sqrt{5}+1}{2}\ket{VH}\right)
\end{equation}
and measure in the following bases:
\begin{equation}\label{Eq:Angle}
\begin{aligned}
	{A_0(\theta)} &= \{\cos(\theta)\ket{H}+\sin(\theta)\ket{V}, \sin(\theta)\ket{H}-\cos(\theta)\ket{V}\},\\
	{A_1(\theta)} &= {A_0(\theta-\pi/4)},\\
	{B_0(\theta)} &= {A_0(\theta+ \pi/2)}, \quad {B_1(\theta)} = {A_1(\theta+\pi/2)}
\end{aligned}
\end{equation}
with $\theta=\arccos\sqrt{1/2+1/\sqrt{5}}\approx13.28^\circ$.

\emph{Experimental setup.---}
We experimentally test the Bell inequalities using human generated random numbers. As shown in Fig.~\ref{Fig:MDLSetup}(b), we first generate a 780 nm pulsed light as pump via a second-harmonic generation (SHG) process. In the experiment, we set the 1560 nm seed laser to be 10 ns pulse width at 100 kHz, then amplified and frequency up-conversion it to a 780 nm pump. The pump is then focused on a periodically poled potassium titanyl phosphate (PPKTP) crystal to create photon pairs at 1560 nm via spontaneous parametric down conversion. The down-converted photon pairs interfere at the polarizing beam splitter (PBS) in a Sagnac based setup \cite{Fedrizzi2007} to create entangled pairs. Non-maximum entangled state can be generated by adjusting the input pump polarization. The entangled pairs are then collected into single mode fibers for detection.

The source lab and the measurement labs are chosen in a straight line. As shown in Fig.~\ref{Fig:MDLSetup}(a), the source is in the middle, and Alice (Bob)'s measurement lab is $87\pm2$ m ($88\pm2$ m) away. The spatial separation makes sure the measurement in Alice's lab will not affect that in Bob's lab, and vice versa. In each pulse period, the random number (if there is one) controls the Pockels cell by applying the zero or half-wave voltage, setting the basis to be $A_0/A_1$ for Alice (or $B_0/B_1$ for Bob). We compensate the polarization drift and align the reference frame using a polarization controller and a half-wave plate.
After the measurement using a Pockels cell and a PBS, the photons are detected with superconducting nanowire single-photon detectors (SNSPD). Note that the total response time of the Pockels cell and the SNSPD is around 150 ns,  less than the ~290 ns separation between source and detection lab.

The random number input was provided by ICFO, who initiated the BIG Bell Test project, and collected and distributed the human generated random numbers. In this project, people play an online game to generate random numbers, which are then transmitted to the experimental labs.  We receive and re-distribute the human generated random numbers with a Python program. A field-programmable gate array (FPGA) board then generates signal and synchronizing pulses based on the real-time random number. We set the system frequency to 100 kHz, the system only works when a batch of random numbers comes in. We record all the detection results and the random numbers using a time-to-digital convertor (TDC) for off-line data analysis.

\emph{Experiment result.---}
In our experiment, we first adjust the pump polarization to be diagonal to generate maximumly entangled state $\ket{\Psi^+}=(\ket{HV}+\ket{VH})/\sqrt{2}$. We measure the visibility in the horizontal/vertical basis as  99.2\% and in the diagonal/anti-diagonal basis as  98.0\%. We further did a state tomography and find that the fidelity to the ideal state is around 97.5\%. By setting the measurement bases according to Eq.~\eqref{Eq:AngleCHSH}, we measure the S value to be 2.804, with $E(A_1,B_1)=-0.751$, $E(A_1,B_2)=0.651$, $E(A_2,B_1)=0.657$, $E(A_2,B_2)=0.745$. Here, the average value is defined by $E(A_x, B_y) = \sum (-1)^{a+b+xy}P(abxy)$.
The classical upper bound of measurement dependent LHVM is given in Eq.~\eqref{Eq:BCHSH}. To achieve the experimentally obtained Bell value with measurement dependent LHVM, we thus have $l<0.1495$. On the other hand, when the input human randomness have $l>0.1495$, the experimentally observed data cannot be explained with LHVMs.

Next, we use the human random numbers to test the MDL inequality. 
The MDL inequality has been previously tested in experiment by Aktas et al. \cite{Aktas15}, but without closing the locality or freedom-of-choice loophole. In our experiment, we closed the locality loophole as shown in Fig.~\ref{Fig:MDLSetup}(a) and the freedom-of-choice loophole with human free will.
In state preparation, we prepare the non-maximum entangled state given in Eq.~\eqref{Eq:Bellinequality}, by adjusting the pump laser to $\cos(69.1^\circ)\ket{H}+\sin(69.1^\circ)\ket{V}$. We perform a state tomography to characterize the produced state and calculate the fidelity to be 98.7\%.

Finally, we analysis the record data to count the coincidence events and calculate the probabilities $P(abxy)$. The result is summarized in Table \ref{tab:MDLhuman}. All experimental trials performed between 30th Nov. and 1st Dec., 2016 are recorded, when the public helped us to generate random numbers. The experimental results are divided into several sections for statistical analysis, with each section including 1 hour of data. We measure the four probabilities $P(abxy)$ and calculate the MDL Bell inequality in Eq.~\eqref{Eq:MDLin} for a given $l$. As this inequality cannot be violated with LHVM satisfying $l\le P(xy|\lambda)$, we calculate the smallest possible value of $l$ such that the equal sign is saturated. When the input randomness has a smaller value of $l$, the observed result cannot be explained with LVHMs.

In experiment, we obtain $l=0.10\pm0.05$ for the MDL inequality by using human random numbers. As a comparison, we use quantum random number generators to choose the basis. With the rest of the setup remaining the same, we obtain $l=0.106\pm 0.007$ which gives a similar $l$ value although with less fluctuation. This is because it is easier to accumulate more data using quantum random numbers than using human random numbers. For human-generated random numbers, we received several thousand bits per second at peak times, but only tens of bits per second at idle times. For quantum random numbers, we have 100 kHz steady random numbers controlling the basis. In total, we have accepted and used around 80 Mbits in human random number based MDL inequality test in two days. In comparison, we test the inequality with quantum random numbers in around 1.5 hours, consuming in the process more than 500 Mbits. 

\begin{table}[htb]
\centering
  \caption{Experimental values of the MDL inequality measurement using human random numbers. MDl value $l=0.10\pm0.05$ is obtained by calculating all the data. We measure in the basis $A_0 B_0$, $A_0 B_1$, $A_1 B_0$, and $A_1 B_1$ to get the probability of $P(0000)$,  $P(0101)$,   $P(1010)$, and  $P(0011)$ for analysis in Eq.~\eqref{Eq:MDLin}. In the probability $P(abxy)$, $xy$ denotes Alice's and Bob's basis, and $ab$ denotes the output bits Alice and Bob count for coincidence. The column "Total" refers to the total coincidence count for all Alice's and Bob's outputs, and "Measured" refers to the coincidence count of the specific outputs listed in $P(abxy)$. The result sums all the experiment trials between 30th Nov. and 1st Dec., 2016, the statistical fluctuation is calculated using 1 hour as a integration time.}
\begin{tabular}{cccc}
\hline
Basis & Measured & Total & $P(abxy)$ \\
\hline
$A_0 B_0$ & 2833 & 34408 & $P(0000)= 0.02093\pm 0.01099 $ \\
$A_0 B_1$ & 100  & 40085 & $P(0101)= 0.00074\pm 0.00048 $ \\
$A_1 B_0$ & 193  & 41009 & $P(1010)= 0.00143\pm 0.00076 $ \\
$A_1 B_1$ & 86   & 19853 & $P(0011)= 0.00064\pm 0.00046 $ \\
\hline
\end{tabular}
\label{tab:MDLhuman}
\end{table}

\begin{table}[htb]
\centering
  \caption{Experimental values of the MDL inequality measurement using quantum random numbers. MDl value $l=0.106\pm0.007$ is obtained calculating all the data. The columns are same as in Table \ref{tab:MDLhuman}. The experiment is done after the test with human random numbers, using the same setup, only substitute the random numbers to quantum random numbers. the statistical fluctuation is calculated using 5 minutes as a integration time.}
  \begin{tabular}{cccc}
\hline
Basis & Measured & Total & $P(abxy)$ \\
\hline
$A_0 B_0$ & 38911  & 463901 & $P(0000)= 0.02101\pm 0.00062 $ \\
$A_0 B_1$ & 1214   & 453939 & $P(0101)= 0.00066\pm 0.00009 $ \\
$A_1 B_0$ & 3246   & 471152 & $P(1010)= 0.00175\pm 0.00023 $ \\
$A_1 B_1$ & 1577   & 462591 & $P(0011)= 0.00085\pm 0.00007 $ \\
\hline
\end{tabular}
\label{tab:MDLqrng}
\end{table}

\emph{Discussion.---}
In this work, we realized a Bell test with the assistance of human free will in order to close the freedom-of-choice loophole. We test the CHSH and MDL inequalities in an experiment that closes the locality loophole.
Comparing the results of the two inequalities, we can see that the MDL inequality tolerates more imperfection of the input randomness, in the sense of influence by the hidden variables in the LHVM. 
Future works could extend the results to randomness amplification that amplifies imperfect human randomness to almost uniform randomness.

\emph{Acknowledgement.}---We thank ICFO for organizing the BIG Bell Test, and the Bellsters for providing random number inputs. We acknowledge L.-F. Y and Q. Fan for outreach in China. This work is briefly reported as part of the Big Bell Test \cite{BBT2018}. This work has been supported by the National Fundamental Research Program (under Grant No. 2013CB336800), the National Natural Science Foundation of China, and the Chinese Academy of Science.

\bibliography{BibBBT}

\end{document}